\documentstyle[aps,prb,epsf,floats]{revtex}

\begin{document}
\draft

\wideabs{
\title{Charge-Stripe Ordering From Local Octahedral Tilts: Underdoped and 
Superconducting $\bf La_{2-x}Sr_xCuO_4$ (0$\le$x$\le$0.30) 
 }

\author{E. S. Bo\v zin, S. J. L. Billinge,
}
\address{Department of Physics and Astronomy and Center for
Fundamental Materials Research,}
\address{Michigan State University, East Lansing, MI 48824-1116.}
\author{G. H. Kwei,
}
\address{Los Alamos National Laboratory, Los Alamos, NM 87545.}
\author{H. Takagi}
\address{Institute for Solid State Physics, University of Tokyo,
7-22-1, Roppongi, Minato-ku, Tokyo, 106, JAPAN
}
\date{\today}

\begin{center}
To appear in {\it Physical Review {\bf B}}
\end{center}

\maketitle

\begin{abstract}
The local structure of La$_{2-x}$Sr$_x$CuO$_{4}$, for $0\le x\le 0.30$,
has been investigated using the atomic pair distribution 
function (PDF) analysis
of neutron powder diffraction data. The local octahedral tilts are
studied to look for evidence of {\rm $[110]$} symmetry (i.e., LTT-symmetry) 
tilts locally, even though the average tilts have {\rm $[010]$} symmetry 
(i.e., LTO-symmetry) in these compounds.  We  argue that this observation
would suggest the presence of local 
charge-stripe order.  We show that the tilts are locally LTO in the
undoped phase, in agreement with the average crystal structure. At
non-zero doping the PDF data are consistent with the presence of local tilt
disorder in the form of a mixture
of LTO and LTT local tilt directions and a distribution of 
local tilt magnitudes.  We 
present topological tilt models which qualitatively explain the origin
of tilt disorder in the presence of charge stripes and show that the
PDF data are well explained by such a mixture of locally small and large
amplitude tilts.

\end{abstract}
\pacs{61.12-q,71.38.+i,74.20.Mn,74.72.Dn,74.72.-h}
}

\section{
Introduction
}
There is now considerable experimental evidence for the existence of 
striped charge distributions in the strongly correlated materials
La$_{2-x}$Sr$_x$NiO$_{4+\delta},
$\cite{tranq;prl94,sacha;prb95,vigli;prb97,tranq;prl97ii} 
La$_{2-x-y}$Nd$_y$Sr$_x$CuO$_{4+\delta},
$\cite{tranq;n95,tranq;prb96,zimme;epl98}
and La$_{1-x}$Ca$_x$MnO$_{3}$.\cite{chen;prl96,chen;jap97,ramir;prl96ii}  
There is also growing
circumstantial evidence that similar striped charge distributions
are present more widely in other high-temperature superconducting
cuprates,\cite{hamme;prb98,egami;js97,statt;prb95,baber;cm97} 
and their possible 
importance in producing the high-temperature superconductivity 
itself is the subject of intensive 
investigation.\cite{emery;prb97,salko;prl96,kroto;prb97,marki;unpub,phill;pnas97,bianc;ssc97} 
A number of recent reviews describe observations of lattice effects
in the high-temperature superconductors in 
general.\cite{egami;b;pphtsv96,egami;pms94,salje;pbhts95,%
mulle;b;pscs93,sigmu;c;psics94,barya;c;lehts92}
However, there is 
currently no direct evidence that these stripe phases exist in 
superconducting samples, although there has been an 
observation where stripes and superconductivity were observed
in the same samples.\cite{tranq;prl97i}
It is important to show that local charge-stripe ordered domains
really exist in the superconducting regions of samples to establish
the importance of charge-stripes in the superconducting phenomenon.

A local structural probe is useful for investigating the existence or
absence of short-range ordered stripes since it is known that the 
charge-stripes produce a structural distortion.  When the stripes are
long-range ordered the structural modulation becomes periodic and
a superlattice is observed in neutron, x-ray and electron 
diffraction.\cite{tranq;n95,tranq;prb96,chen;prl96,chen;jap97,%
tranq;prl97i,saini;prb97}
The absence of superlattice peaks can mean that the stripe order has
disappeared.  However, it is also possible that the stripes persist
locally but are not long-range ordered and are fluctuating.  In this
latter case the {\it local} structural distortion due to the 
inhomogeneous
charge distribution will persist even when the superlattice peaks have
disappeared. A probe of local structure can therefore 
give information about whether local stripe order exists even in the 
absence 
of superlattice peaks.  In this study we use the atomic pair
distribution function (PDF) analysis  \cite{warre;bk90,%
egami;mt90,billi;b;lsfd98} of powder neutron 
diffraction data to look for evidence of local charge stripe order in 
La$_{2-x}$Sr$_x$CuO$_{4+\delta}$.

An amenable system for this
kind of  study is the series of compounds based around 
La$_{2-x}$(Sr,Ba)$_x$CuO$_4$ (2:1:4 compounds) because in these 
compounds 
collective tilts of the CuO$_6$ octahedra exist which couple strongly 
to the electronic 
system\cite{axe;prl89,crawf;prb91,buchn;pc94,buchn;prl94,billi;prl93} 
and the charge-stripes.\cite{tranq;n95,tranq;prb96}
In particular, it has been observed that long-range ordered 
charge-stripes
are only seen in the cuprate system when the octahedra tilt 
collectively 
about axes along the Cu-O bonds, i.e., in the [110] crystallographic 
directions
in the standard crystallographic setting 
(P4$_2$/ncm for the LTT phase).
\cite{tranq;n95,tranq;prb96,tranq;prl97i}  
This tilt pattern is the 
one observed in the low temperature tetragonal (LTT) 
phase.\cite{cox;mrssp89}  
Most of the superconducting compositions are in the alternative low 
temperature orthorhombic (LTO) phase\cite{yvon;zpb89} which 
has the octahedra tilting {\it on average} about axes parallel to the 
[100] crystallographic direction. It is thought that since the 
charge-stripes lie in [110] directions in the 
lattice (i.e., along Cu-O bonds), the [100] type LTO tilts prevent
the stripes from ordering statically over long range.  Because the 
charge-stripes are strongly coupled to the octahedral 
tilts,\cite{tranq;n95,tranq;prb96} we can use 
the octahedral tilts as a probe of the local stripe order.

In a previous paper,\cite{billi;prl94} we showed that in 
La$_{1.875}$Ba$_{0.125}$CuO$_{4}$, local regions of LTT-type octahedral
tilt order persist in the LTO phase above the LTO-LTT phase transition.
It was suggested that the LTO phase is made up of short-range ordered 
LTT-domains but that a linear combination of different LTT variants
resulted in the average LTO tilt order observed in the 
crystallographic structure.  There also are other observations that
the local tilts in 2:1:4 materials are not always the same as those
measured 
crystallographically.
\cite{hamme;prb98,egami;rsss87,billi;jpcs96,haske;prl96,haske;prb97,hamme;prl93}
At the time of the earlier paper,\cite{billi;prl94} 
charge stripes had not been 
observed in the cuprates and it was unclear why
the LTO phase should be made up of inhomogeneous domains of 
short-range
LTT-order.~\cite{marki;pc92i,marki;jpcs97}  
However, the presence of dynamic charge-stripe 
domains provides
a very natural explanation of this unexpected result.  

In the present work we have studied the local octahedral tilts in
La$_{2-x}$Sr$_x$CuO$_{4}$ to look for evidence of local LTT symmetry
tilts. We chose to study the Sr doped, rather than the Ba doped, 
systems 
since ionic size effect disorder is less in this system. This is due 
to the 
fact that Sr$^{2+}$ is closer in size to La$^{3+}$ than is Ba$^{2+}$. 
This means that the disorder introduced into the structure due to
ionic size effects is minimal and any disorder in the local tilts will
originate mainly from an inhomogeneous electronic charge distribution.

In this paper we show that the local octahedral tilt amplitude can
be measured accurately using PDF techniques.  If the tilt amplitude 
is large
enough it is also possible to distinguish the tilt direction.  We show
that the local tilts are clearly LTO in nature in the undoped compound,
as expected. However, away from zero doping ($x>0$) the PDF is 
consistent with
the presence of tilt disorder in the form of a mixture of tilt directions
and a mixture of tilt amplitudes.  As $x$ increases at low temperature
the local tilt amplitude decreases smoothly following the behavior of the 
average tilts. However, small but finite tilts persist locally when the
sample goes into the high temperature tetragonal (HTT) phase above $x=0.20$.  
In the Discussion section we present topological tilt models which 
qualitatively explain the origin of tilt disorder in the presence of 
charge stripes and show that the PDF data are well explained by a mixture 
of locally small and large amplitude tilts.

\section{
Experiment
}
Powder samples were prepared by conventional solid state 
reactions. Mixtures of La$_2$O$_3$, SrCO$_3$, and CuO were calcined 
at various 
temperatures between $900^{\circ}C$ and $1050^{\circ}C$ with several 
intermediate grindings. The products were sintered at $1100^{\circ}C$ 
for 100 h, followed by an oxygen anneal at  $800^{\circ}C$ for 100 hours.

Time-of-flight neutron diffraction measurements were carried out
on La$_{2-x}$Sr$_x$CuO$_4$ (LSCO) powder samples (approximately 10 g 
of each) 
for the range of doping ($0\leq x \leq 0.30$) at 10~K temperature. 
Experiments were performed on the Special Environment Powder 
Diffractometer 
(SEPD) at the Intense Pulsed Neutron Source (IPNS) at Argonne National 
Laboratory.

The data were corrected for experimental 
effects and normalized\cite{billi;b;lsfd98} to obtain  the
total scattering function $S(Q),$\cite{warre;bk90} where $Q=\arrowvert 
{\bf Q}\arrowvert =\arrowvert {\bf k}-{\bf k_0}\arrowvert $ represents 
the momentum transfer magnitude for the scattering. 
The PDF function, $G(r)$, is then obtained by a Fourier 
transformation according to, 
\begin{equation}
G(r)= \frac{2}{\pi }\int_{0}^{\infty} Q [S(Q)-1] \sin (Qr)\> dQ.
\end{equation}
The function $G(r)$ 
gives the probability of finding an 
atom at a distance $r$ from another atom. This function is obtained 
directly from the Fourier transform of the neutron-diffraction data 
and is used to investigate 
features of the {\it local} structure of a material.
\cite{billi;b;lsfd98} 

Local structural information is obtained from the data by a process
of modeling.\cite{billi;b;lsfd98} A model structure is proposed and the 
PDF calculated.
Parameters in the model are then varied to optimize the agreement
between the calculated and measured PDFs.  The procedure is
similar to the traditional reciprocal-space based 
Rietveld method\cite{rietv;jac69} for obtaining crystal 
structures from
powder diffraction data; however, it is carried out in real-space and
yields the local structure rather than the average crystal 
structure.\cite{billi;b;lsfd98}
From the refinements, different 
structural information can be extracted, such as lattice parameters, 
average atomic positions and amplitudes of their thermal motion, 
atomic displacements, and magnitudes of local octahedral tilts.
The modeling procedure has been described in detail 
elsewhere.\cite{billi;b;lsfd98}

\section{
Results
}

\subsection{Local Tilt Amplitudes as a Function of Doping}

\begin{figure}[tb]
\begin{center}$\,$
\epsfxsize=2.8in
\epsfbox{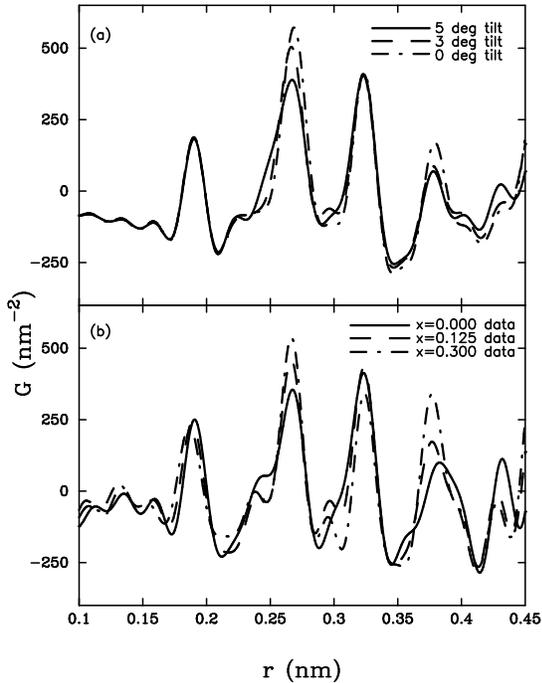}
\end{center}
\caption{(a) Comparison between three {\it model} PDFs. 
The models have LTO symmetry with
5$^{\circ}$ tilts (solid line), 3$^{\circ}$ tilts (dashed line), 
and 0$^{\circ}$ tilts (dash-dotted line). (b) Three different 10~K 
{\it data} PDFs that approximately correspond to the same tilt 
magnitudes as shown in the models in panel (a): 
$x=0$ (solid line, $\approx 5^{\circ}$ tilt), $x=0.125$ 
(dashed line, $\approx 3^{\circ}$ tilt), and $x=0.30$ 
(dash-dotted line, $\approx 1^{\circ}$ 
tilt). The PDF technique clearly differentiates between the presence 
and absence of the tilts and these qualitative differences are evident in
the data.}
\protect\label{fig;fig1}
\end{figure}

Initially we investigate the appearance in the PDF of the octahedral 
tilts and show the extent to which the PDF technique can 
detect their presence. In Fig.~\ref{fig;fig1}(a), we compare three 
{\it model} PDFs to show qualitatively the effect of a change in the 
local octahedral tilt amplitude on $G(r$).
The PDFs are shown for the LTO structure with three different magnitudes 
of octahedral tilt: 0$^\circ$ (dash-dotted 
line), 3$^\circ$ (dashed line) and 5$^\circ$ (solid line). 
The initial parameters used in the models were obtained by converging the
fit of the LTO model-PDF to the data-PDF from the $x=0.125$ data-set at 
10~K,
using the RESPAR Real-Space Rietveld program.\cite{billi;b;lsfd98}  
The octahedral tilt angles were then
calculated independently from the positions of the in-plane oxygen (O1)
and the out-of-plane oxygen (O2). These tilt angles are denoted by
$|\theta_{O1}|$ and $|\theta_{O2}|$ respectively.
The tilt angle $|\theta_{O1}|$ was 
obtained from the $z$-displacement of O1 using
\begin{equation}
|\theta_{O1}| = \left|\arctan\left(
{c\,\delta O1_z \over b\,\delta O1_y}\right)\right|,
\protect\label{eq;thetao1}
\end{equation}
where $\delta O1_z$ and $\delta O1_y$ are the $z$ and $y$ fractional 
coordinates 
respectively of O1, and $c$ and $b$ are the corresponding lattice 
parameters.
Similarly, the average tilt is obtained from displacements of O2 using
\begin{equation}
|\theta_{O2}| = 
\left|\arctan\left({b\,\delta O2_y \over c\,\delta O2_z}\right)\right|.
\protect\label{eq;thetao2}
\end{equation}
In the case where the octahedral tilts are rigid, 
$|\theta_{O1}|=|\theta_{O2}|$.  The tilt angles were then artificially
adjusted to 
5$^\circ$, 3$^\circ$ and 0$^\circ$ in the model
and the atomic positions for O1 and O2
determined from Eqs.~\ref{eq;thetao1} and \ref{eq;thetao2}.  All other
parameters in the model were held constant.  In this way we could compare
the effect of a change in tilt magnitude on the PDF neglecting other changes
such as changes in bond-length or lattice parameter.
As expected, the Cu-O nearest neighbor peak at 1.89~\AA\ is unaffected
(the tilts are essentially rigid) but the peak at 2.7~\AA , which contains
the La-O1 and La-O2 correlations (as well as the O1-O1 correlations),  
shows a particularly large change.  

The lower panel, Fig.~\ref{fig;fig1}(b), presents PDFs obtained from the 
experimental data: $x=0$ doping (solid line), where tilts of 
approximately 5$^{\circ}$ are present, $x=0.125$ doping 
(dashed line) with tilts of approximately 3$^{\circ}$, and
$x=0.30$ doping (dash-dotted line), with average tilts of zero. 
The changes in the data with doping
are large and qualitatively similar to those expected for reductions in
local tilt amplitude.  In addition, a shortening of the average Cu-O bond
is evident by a shift to the left of the nearest neighbor Cu-O peak, as 
expected from the average structure.  This also gives an indication of the
sensitivity of $G(r)$ to small structural changes such as this lattice 
contraction.

\begin{figure}[tb]
\begin{center}$\,$
\epsfxsize=2.8in
\epsfbox{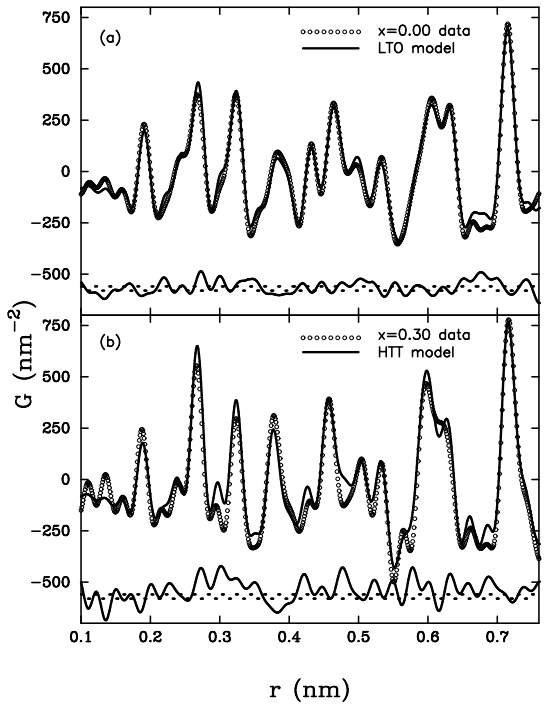}
\end{center}
\caption{(a) Fully converged PDF for the LTO model (solid line), 
and experimentally obtained PDF for LSCO with x=0 at 10~K (open circles). 
The difference curve is plotted below as a solid line.  The dotted line
shows the expected errors at the level of two standard deviations.
(b) Fully converged PDF for the HTT model (solid line),  
and LSCO with x=0.30 at 10~K (open circles).}
\protect\label{fig;fig2}
\end{figure}
\begin{figure}[tb]
\begin{center}$\,$
\epsfxsize=2.8in
\epsfbox{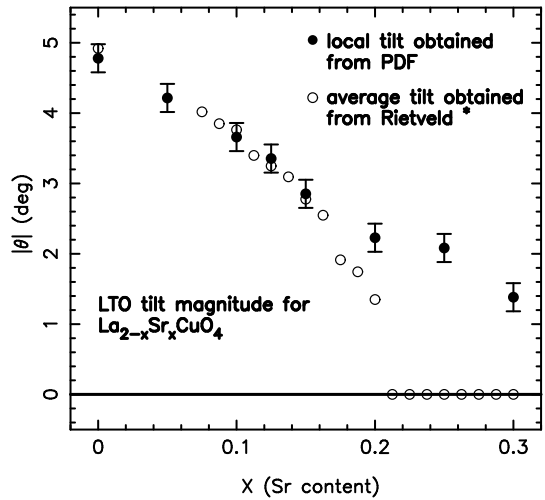}
\end{center}
\caption{Dependence of the local octahedral tilt angle magnitude 
(black circles), $|\theta|$, on Sr content, $x$. Open circles present average 
tilt angle magnitude obtained from Rietveld refinement done by Radaelli and 
collaborators (see text). The result shows that $|\theta|$ smoothly decreases 
when x is increased. However, significant local tilts persist even when the 
average tilts disappear. The data were collected at 10~K.}
\protect\label{fig;fig3}
\end{figure}

We have determined quantitatively the amplitudes of the 
local octahedral tilts 
as a function of doping using the LTO model and the Real-Space Rietveld
modeling program. 
Characteristic fits are shown in Fig.~\ref{fig;fig2} for the undoped material
(average tilts are large) and the overdoped material 
(average tilts are  zero). Difference curves are plotted below the
PDFs.  The agreement of the HTT model to the $x=0.3$ data is significantly
poorer than the fit of the LTO model to the undoped material.  As we
discuss later, this is because small but finite local tilts are still present
in the structure even in the HTT phase.

The local tilt angles obtained from the fits are shown
in Fig.~\ref{fig;fig3} plotted as a function of doping, 
shown as filled circles.
For comparison, average tilt angles obtained from a crystallographic
analysis~\cite{radae;prb94i} are also plotted (shown as open circles). 
There is excellent agreement for lower dopings.  
The agreement is less good for compositions above $x=0.15$.  In this
region the long-range 
average tilt angle is constrained to be zero because of the
change in the average symmetry accompanying the phase change from LTO
to HTT.  However, in the PDF a better fit to the data is obtained when
the LTO model is used, rather than the HTT model, and a small but finite
($\approx 1-2^\circ$) local tilt is refined.\cite{note1}
The use of the LTO model to
describe the local structure, even when the global structure is clearly
HTT, is justified since the broken symmetry phase may persist locally.
However, this analysis does not imply that, in our samples, 
the global structure is LTO in this region of doping.

\subsection{Local Tilt Directions as a Function of Doping}

\begin{figure}[tb]
\begin{center}$\,$
\epsfxsize=2.8in
\epsfbox{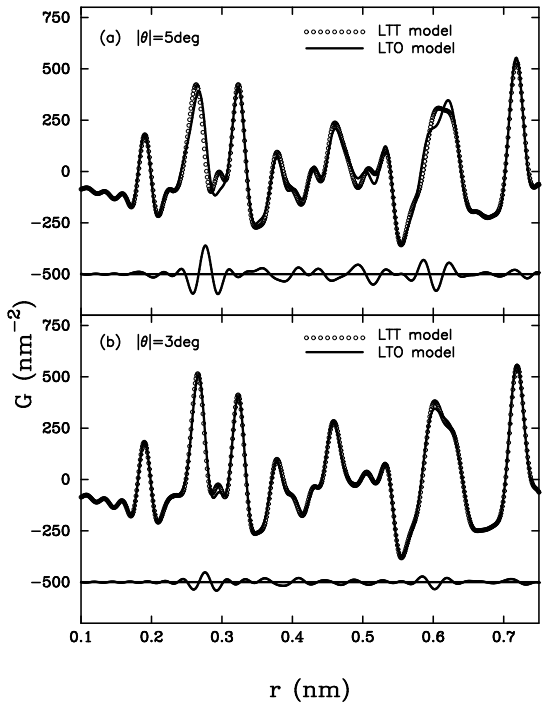}
\end{center}
\caption{Difference in model PDFs for the LTT and LTO tilt symmetry: 
(a) 5$^{\circ}$ case, and (b) 3$^{\circ}$ case. The PDF for the LTT 
model is given as a solid line, while that for the LTO model is presented 
with open circles. Difference curve is given below PDFs for both cases.}
\protect\label{fig;fig4}
\end{figure}
\begin{figure}[tb]
\begin{center}$\,$
\epsfxsize=2.8in
\epsfbox{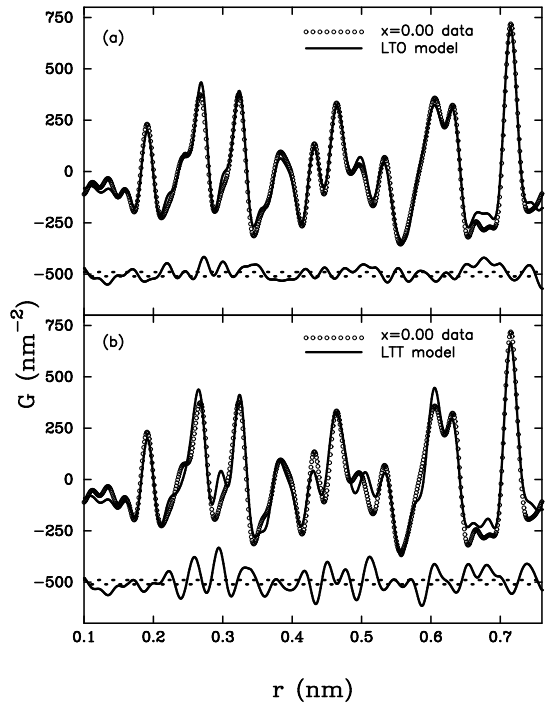}
\end{center}
\caption{(a) Fully converged model PDF with the LTO tilt symmetry (solid line) 
compared to experimental PDF for x=0 at 10~K (open circles). 
Difference curve is shown below the PDFs as a solid line.  The dotted line
shows the expected errors at the level of two standard deviations.
(b) The same for the case of LTT tilt symmetry.}
\protect\label{fig;fig5}
\end{figure}

We would like to use the PDF technique to differentiate between local tilts
with the [010] (LTO: tilt axis 45$^{\circ}$ rotated with 
respect to planar Cu-O bond) and [110] (LTT: tilt axis is parallel to 
the Cu-O bond) tilt symmetries since evidence for
local LTT-like tilts might suggest the presence of charge stripes.

In Fig.~\ref{fig;fig4} we show a comparison of two model PDFs, one 
 where the tilts belong to the LTO type, and the other that is of LTT type. 
In Fig.~\ref{fig;fig4}(a) we show the difference for 
the case of a 5$^{\circ}$ 
tilt magnitude and in Fig.~\ref{fig;fig4}(b) for a smaller
3$^{\circ}$ tilt. It is important to emphasize that the only difference 
between our models is in the tilt symmetry (or directions of 
the tilts), keeping all other parameters constant, including tilt amplitude. 
Therefore, the 
difference in the PDFs that is shown is {\it only} due to the change in
the tilt symmetry.

As is usual with the PDF, some regions of the function are much more strongly
affected by the change than others. A change in tilt direction 
affects the peaks around 2.67~\AA\   and 2.95~\AA, that correspond 
to (La/Sr)-O1 and (La/Sr)-O2 bonds respectively. There is also a large effect
on the peak located close to 6.0~\AA. Elsewhere in the PDF the changes
are small. Notice that the 
peaks at $\approx 2.8$~\AA\ give a characteristic ``W-shaped'' feature in the 
difference curve, while the peak at $\approx 6$~\AA\ gives a characteristic 
``M-shaped'' feature.  The changes are similar, but much smaller, as the
local tilt amplitude diminishes.

In Fig.~\ref{fig;fig5} we present fully converged fits of the LTO and 
LTT models
to the undoped 
material.
It is clear that the undoped material has local octahedral tilts that have 
the LTO symmetry: the fit is much better everywhere with the LTO model, 
and particularly large fluctuations are observed around $r = 2.8$~\AA\
in the LTT model.

\begin{figure}[tb]
\begin{center}$\,$
\epsfxsize=2.8in
\epsfbox{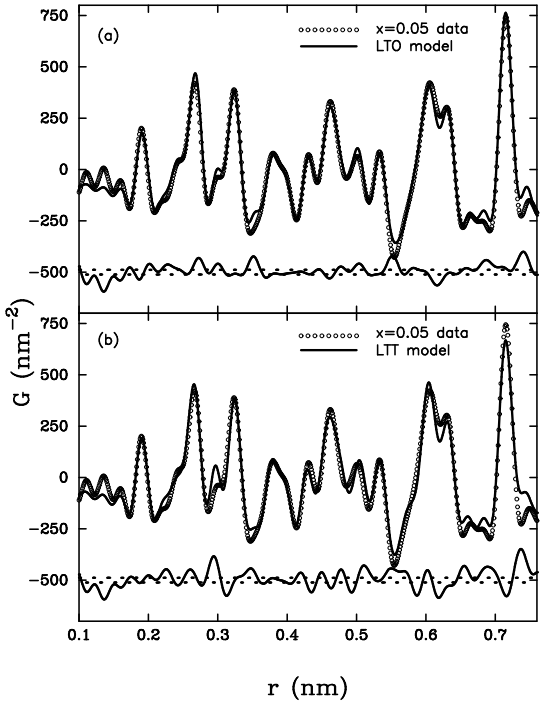}
\end{center}
\caption{(a) Fully converged model PDF with the LTO tilt symmetry (solid line) 
compared to experimental PDF for x=0.05 at 10~K (open circles). 
The difference curve is 
shown below the PDFs as a solid line.  The dotted line
shows the expected errors at the level of two standard deviations.
(b) The same for the case of LTT tilt symmetry.
}
\protect\label{fig;fig6}
\end{figure}
\begin{figure}[tb]
\begin{center}$\,$
\epsfxsize=2.8in
\epsfbox{bozin6987.fig}
\end{center}
\caption{(a) Fully converged model PDF with the LTO tilt symmetry (solid line) 
compared to experimental PDF for x=0.10 at 10~K (open circles). The difference c
urve is 
shown below the PDFs as a solid line.  The dotted line
shows the expected errors at the level of two standard deviations.
(b) The same for the case of LTT tilt symmetry.
}
\protect\label{fig;fig7}
\end{figure}
\begin{figure}[tb]
\begin{center}$\,$
\epsfxsize=2.8in
\epsfbox{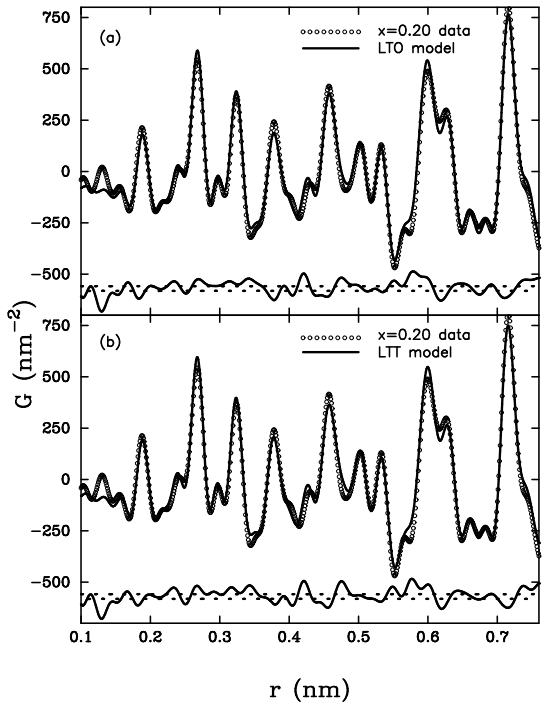}
\end{center}
\caption{(a) Fully converged model PDF with the LTO tilt symmetry (solid line) 
compared to experimental PDF for x=0.20 at 10~K (open circles). 
The difference curve is 
shown below the PDFs as a solid line.  The dotted line
shows the expected errors at the level of two standard deviations.
(b) The same for the case of LTT tilt symmetry.
}
\protect\label{fig;fig8}
\end{figure}

We are interested if evidence for local LTT-like tilts appears as 
the CuO$_2$ planes are doped.
In the 
set of Figures~\ref{fig;fig5}--\ref{fig;fig8} we show the PDFs 
for samples with different dopings: again, upper panels 
show the PDF for fully converged 
LTO models (solid line) and experimental PDFs (open 
circles) with corresponding difference curves; lower panels compare PDFs 
for fully converged LTT models (solid line) with the PDFs obtained from 
the data (open circles), and give corresponding difference curves. 

In Fig.~\ref{fig;fig6}, for $x=0.05$ 
we again see that the LTO fit is better than the LTT one 
overall, and the characteristic $M$-feature is evident in panel (b). 
The underlying local tilts are still predominantly LTO-like.

For $x=0.10$ the situation is less clear. The tilt amplitude is now
down to 3.6$^\circ$ making differentiation between LTO and LTT harder. 
Both models fit quite well.  However, we note that in the critical region
around $r = 2.8$~\AA\ the LTT model actually has a {\it better} agreement 
with the
data than the LTO model and in this case the difference curve from the 
LTO model 
has an $M$-shaped
fluctuation centered at $r = 2.8$~\AA\ consistent with some local LTT-like 
tilts.

As doping is raised beyond it becomes difficult to
differentiate between the LTO and LTT models. Both models 
fit quite well as
evidenced, for example, by the $x=0.20$ data set shown 
in Fig.~\ref{fig;fig8}.
 
\section{
Discussion
}

\subsection{Tilt magnitude vs. doping}

The evolution of the {\it local} tilt magnitude, determined by 
fitting the LTO
model to the PDFs from the data, decreases smoothly with 
increasing doping.
This is consistent with crystallographic studies which show a 
smoothly decreasing {\it average} tilt amplitude 
with doping.~\cite{radae;prb94i,brade;pc94} 
Furthermore, it is clear from Fig.~\ref{fig;fig3}
that the average {\it local} tilt amplitude is quantitatively 
the same as the
long-range ordered average tilt.  Although this may not seem 
surprising,
it is widely observed in these materials that the 
average {\it local} tilt 
amplitude can deviate significantly from the long-range 
ordered value.  This occurs
when finite local tilts exist but the tilt directions become 
disordered 
and the long-range tilt order is not preserved.  This is very 
similar to a 
ferromagnetic-paramagnetic transition where the local moment 
survives
but the long-range ordered moments average to zero.  This behavior 
is
seen in the temperature dependence of the tilts in these 
systems.
\cite{billi;jpcs96,haske;prl96,haske;prb97,bozin;pb98,bozin;unpub98}
The present result shows that the real, local, tilt is decreasing
as $x$ increases:
in these materials the tilts are not just becoming disordered at
low temperature by the action of doping but are smoothly decreasing.

This is easy to understand by the argument that the 
copper-oxygen bond
is shortening as holes are doped and this anti-bonding 
bond is stabilized.
The plane buckling occurs because the CuO$_2$ planes 
are constrained to fit
continuously with the rare-earth oxide charge reservoir 
layers.\cite{zhou;prb97}
  The buckling allows them to maintain a continuous 
structure with the charge 
reservoir layers while relieving stress introduced 
because the Cu-O bonds 
are just too long to fit perfectly.  As the Cu-O bonds 
are shortened by doping
the amplitude of tilt needed to relieve this stress is 
reduced and the
tilts smoothly decrease.

In the high-doped region there is not perfect agreement between the 
local and long-range ordered tilt amplitudes. 
The crystallographic result is constrained to be zero above
$x=0.2$ where the global structure goes to the HTT phase.  
There is a remanent amount of tilting of around
1-2$^\circ$ which persists
even into the HTT phase.
This is not completely unexpected because of the presence 
of Sr impurities
in the structure which would be expected to modify the 
octahedral tilting
in their immediate vicinity.\cite{hamme;prb98,haske;prb97}  
It does not necessarily imply any significance
with respect to the charge doping in the planes themselves.

\subsection{Tilt direction vs. doping}

If charge is doped into the material, it can be either distributed 
uniformly 
or it can be localized and over some doping range it can 
form charge stripes. 
Localized charge would cause local polaronic 
distortions, forcing the octahedra to change their shape, and
affecting the magnitude of the tilts in the vicinity of the polaron.  
In addition,
the presence of charge stripes should affect the
direction of the octahedral tilting since it has been observed that
long-range ordered charge stripes in the 2:1:4 materials have 
only been seen in samples in the LTT phase.
Thus, the observation of {\it local} LTT-like tilts would be 
evidence for the
presence of {\it local} charge-stripe order.  

Local LTT-like tilts have been seen in the LTO phase of 
La$_{1.875}$Ba$_{0.125}$CuO$_4$.\cite{billi;prl94} 
In this case tilt amplitude is larger than it is in 
the Sr doped case 
reported here (3.8$^\circ$ for Ba,~\cite{bozin;pb98} 
x=0.15, vs 2.9$^\circ$ 
for Sr, x=0.15, both at 10~K) which explains why it 
was more clearly evident 
in the Ba doped case.  
 At the time, there
was no obvious physical justification for the presence 
of local LTT-domains
which were globally disordered.  However, it could be 
explained if
 the local LTT-like tilts are
stabilized by the presence of local charge stripe order 
and if the charges
are fluctuating dynamically.  Long-range LTT tilt order 
would also be
inhibited at doping fractions away from rational
numbers where the holes can order commensurately with the 
lattice, even if 
the tilts were locally LTT-like. Models proposed in the 
earlier work
showed that global LTO symmetry can be recovered by a 
linear superposition
of two degenerate local LTT variants.  Thus, spatially 
or temporally fluctuating
LTT domains can yield a global LTO tilt symmetry.

The picture of local, fluctuating, LTT-like domains
is also consistent with recent thermal conductivity 
measurements~\cite{baber;cm97} where the apparently 
contradictory result was obtained that the thermal 
conductivity ($\kappa$) of insulating samples of rare-earth 
doped La$_{2-x}$Sr$_{x}$CuO$_4$ 
 was {\it higher} than that of similar samples which were
conducting.  Clearly the contribution of the charge carriers 
to the thermal
transport is negligible, which is not surprising because of 
the low carrier 
density; but also implied by this result is that the 
inelastic phonon scattering is significantly greater in 
metallic samples 
than in insulating ones.  The authors proposed that 
scattering was occurring
off local domains of stripes.  We would like to add that, 
in the present picture,
this phonon scattering would be greatly enhanced if 
there were fluctuating
domains of {\it tilt} disorder associated with 
these striped domains.
The result that there are regions of locally fluctuating LTT
tilts is also consistent with measurements of the anelastic 
spectra from La$_2$CuO$_{4+\delta}$~\cite{corde;prb98} which 
indicate elastic relaxations consistent with octahedral tilts
tunneling between different LTT domains.

The first result we have demonstrated here is 
that the local tilt direction
in undoped La$_2$CuO$_4$ is [010] (LTO), 
and that the PDF is clearly 
sensitive to the tilt direction for tilts 
of magnitude $\approx 5^\circ$.
The local tilt direction and magnitude is the same as the average
crystallographic 
tilt direction and magnitude for the undoped material at low
temperature and there is no tilt disorder.
\cite{radae;prb94i,brade;pc94}

For doping levels greater than zero the PDF results 
are suggestive of the 
presence of local LTT-like tilts. 
The tilt directions and magnitudes are disordered, 
and also could be 
fluctuating and could originate from an inhomogeneous 
charge distribution 
in the CuO$_2$ planes. 
  As we discuss below, we expect that charge localization in a 
background of octahedral tilts will give rise to complex 
patterns of tilt
disorder which our simplistic modeling using pure LTO and LTT models 
cannot hope to reproduce. Below we describe qualitatively
different models of tilt defects which give insight as to when LTO and
LTT tilts, respectively, are stabilized.

\subsection{Tilt Defect Models}

We introduce here  several different models for the tilt distortions
which might be expected as a result of the presence of localized charges
in the tilt background.  Our starting point is based on the two following
observations: (a) Buckling appears because Cu-O bonds are too long to 
match the bonding in the La-O intergrowth layers,\cite{zhou;prb97} and 
(b) Cu-O bonds become shorter on doping.

In the absence of holes the equilibrium tilt amplitude is 5$^\circ$.
On average, the tilts disappear for $x\approx 0.2$ which corresponds to
a nominal copper charge-state of +2.2.  If charges completely localize,
the charge state of the copper is +3 which is easily high enough to  
remove the
local tilt.  However, between the localized charges the material 
is essentially
undoped and large tilts are expected.  Because the CuO$_2$ planes are 
a continuous network of corner shared octahedra, these untilted defects
will introduce strains and there is expected to be a distribution of
tilt amplitudes. However, it is possible to show
geometrically that local LTT tilted octahedra (LTT-defects) can help 
reduce the strains.

We make the following assumptions in our models. First, the undistorted
tilt pattern is LTO-like.  Second, that the localized doped holes form
into stripes with one hole associated with every second CuO$_6$ octahedron
along the stripe.\cite{tranq;n95}  
The nominal doping then determines the average separation
of neighboring stripes.  We then considered two possibilities.  The first
is that the charge essentially delocalizes along the chain making the 
Cu-O bonds along the chain short, but not shortening the Cu-O bonds 
perpendicular to the stripe.  The second is that each charge localizes on
a single CuO$_6$ octahedron, similar to a Zhang-Rice 
singlet,\cite{zhang;prb88}
 sharing charge density equally between each of the four 
in-plane Cu-O bonds.  
We introduce
these defects into the background of LTO tilts with the constraint that
the octahedra are corner shared and displacements of an oxygen ion must
be the same for neighboring octahedra.  This gives rise to 
extended defects
which we can identify using only these topological constraints.
We will
refer to the first case as Model~I and the second case as Model~II.
Both models were made for the case of $x=0.125$ doping.

\subsubsection{Model I}

\begin{figure}[tb]
\begin{center}$\,$
\epsfxsize=2.8in
\epsfbox{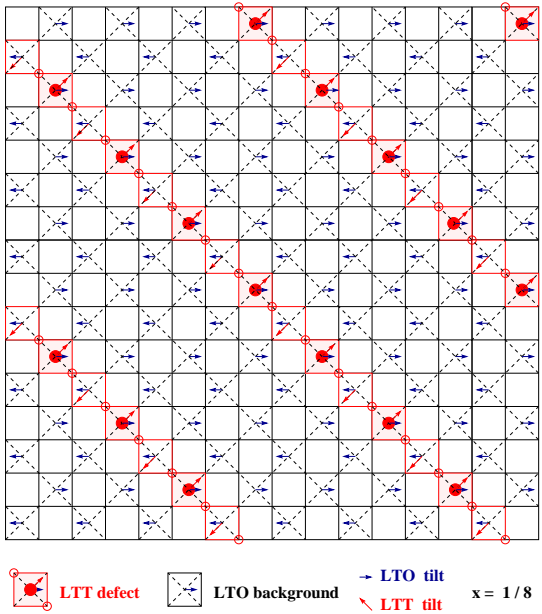}
\end{center}
\caption{Schematic view of the Model~I (see text) 
tilt pattern in the CuO$_2$ plane
in the presence of charge stripes.  Corner shared CuO$_6$ octahedra
are denoted by squares with dashed crosses inside.  The displacement of
the apical oxygen above the plane due to octahedral tilting 
is shown with small arrows. In-plane O1 ions lie at the corners of the
octahedra and are displaced up or down by the tilts (not shown). Open 
circles at the corners indicates an O1 ion which is undisplaced and lies in
the plane.  The presence of a localized hole is indicated by a black circle.}
\protect\label{fig;fig9}
\end{figure}

With the LTO tilts, all of the O1 ions are displaced 
out of the plane, either
up or down.  By shortening two of the in-plane bonds along the 
direction of the charge stripe we expect this to 
bring the two O1 ions which lie along the stripe (denoted by open 
circles in Fig.~\ref{fig;fig9}) into the plane from 
their displaced positions.  
The bonds perpendicular to the stripe remain long. 
In this case, the local tilt
persists, but changes from LTO-like to LTT-like 
in the stripe.  The LTO
tilt pattern between the stripes is preserved, 
though the amplitude of the
LTO tilts may be locally distorted 
where it joins the LTT-stripes.

From 
Fig.~\ref{fig;fig9}, for ${1\over 8}$~doping we expect to get 
a ratio of LTT:LTO tilted octahedra of 2:6. 
This number is given by the ratio of the width 
of the stripe to the
width of the undoped domain; in this case 1:3. 
In general, for 
${1\over n}$ holes per copper this ratio is 
given by 1:${{n-2}\over 2}$.  

This model predicts that in the presence of charge 
stripes the local tilt
distribution should be a mixture of LTO and 
LTT tilts with a majority of
LTO tilts in the doping range up to $x=0.25$.  

\subsubsection{Model II}

\begin{figure}[tb]
\begin{center}$\,$
\epsfxsize=2.8in
\epsfbox{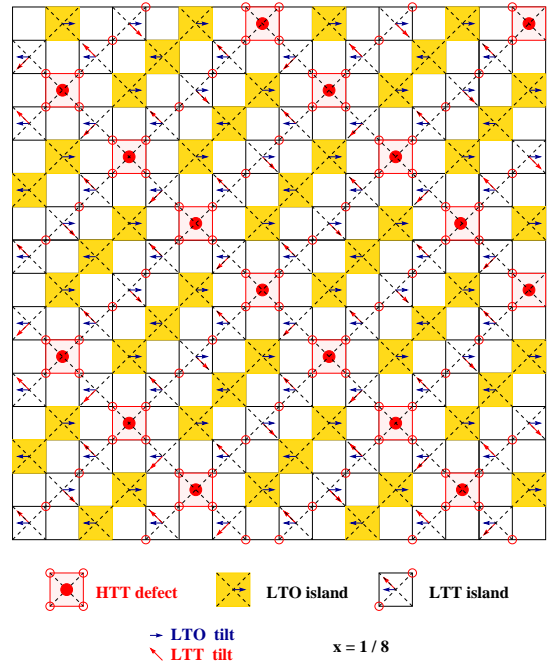}
\end{center}
\caption{Schematic view of the Model~II (see text) 
tilt pattern in the CuO$_2$ plane
in the presence of charge stripes.  Corner shared CuO$_6$ octahedra
are denoted by squares with dashed crosses inside.  The displacement of
the apical oxygen above the plane due to octahedral tilting 
is shown with small arrows. In-plane O1 ions lie at the corners of the
octahedra and are displaced up or down by the tilts (not shown). Open 
circles at the corners indicates an O1 ion which is undisplaced and lies in
the plane.The presence of a localized hole is indicated by a black circle at
the center of the octahedron.}
\protect\label{fig;fig10}
\end{figure}

In this model the localized holes on every second 
site along the stripe
produce octahedra which are completely
untilted.  This is equivalent to placing an 
HTT defect in the LTO background.
Because the out-of-plane displacements of the 
O1 ions are removed, neighboring
octahedra both along the direction of 
the stripe, and perpendicular to the
stripe, take on an LTT-like tilt, 
as shown in Fig.~\ref{fig;fig10}.
The tilt axes of the LTT-like tilts 
propagating parallel and
perpendicular to the stripes are rotated by 90$^\circ$
with respect to each other.  As we pointed out in our earlier 
paper,\cite{billi;prl94} the linear superposition of these two 
degenerate LTT-variants yields the LTO symmetry 
of tilts on average.

As is clear from Fig.~\ref{fig;fig10}, patches of 
LTO-like tilts also 
persist in this model; however, the extent of the 
LTT-like tilting 
is greater than in Model~I.  In the case of $x={1\over 8}$ 
the ratio of
HTT:LTT:LTO is 1:4:3.  In general, for ${1\over n}$ holes 
the ratio will
be 
\begin{equation}
1:{n\over 2}:{n-2\over 2}.  
\label{eq;modii}\end{equation}
One assumption inherent in this model is 
that the LTT-stripes 
propagating perpendicular to the charge-stripes 
persist all the way to the
neighboring stripe.  This is an approximation 
since the Cu-O bonds are
long along this stripe and O1 ion displacements
 may not completely disappear along the stripe 
(even though these ions are
depicted as open circles in Fig.~\ref{fig;fig10}).  
Thus, when the charge-stripes
are well separated at low doping we might expect that the ratio of 
LTT:LTO is smaller than predicted by Eq.~\ref{eq;modii}.

\subsubsection{Discussion}

\begin{figure}[tb]
\begin{center}$\,$
\epsfxsize=2.8in
\epsfbox{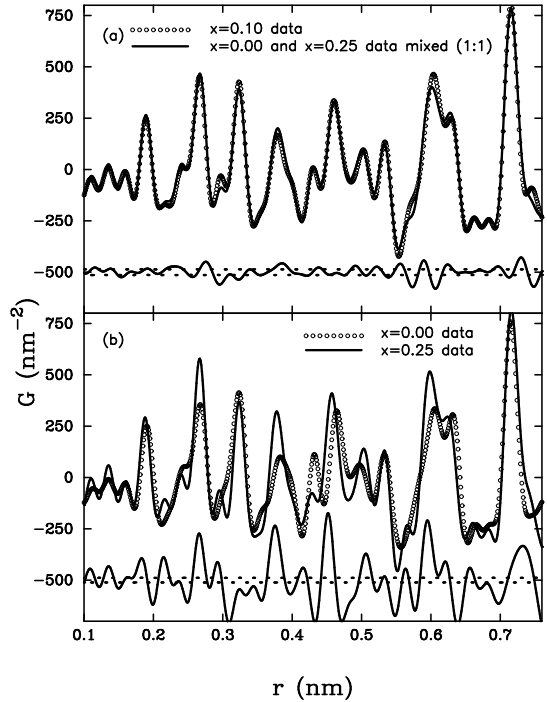}
\end{center}
\caption{(a) Comparison of PDFs produced by a mixture of large and small tilt
amplitudes and the data from the $x=0.10$ sample. The solid line is the 
PDF obtained by mixing the $x=0$ data set (5$^\circ$ tilts) with the 
$x=0.25$ ($<2^\circ$ tilts) data set in the ratio 1:1 to mimic the effect of 
a coexistence of large and small tilts in the local structure.  The open
circles show the PDF from the the $x=0.10$ sample. (b) The $x=0$ and
$x=0.25$ data-PDFs are plotted for comparison.  There are large differences
between these PDFs, yet when they are mixed they reproduce the PDF of the
intermediate composition extremely well. The data were collected at 10~K.}
\protect\label{fig;fig11}
\end{figure}

These tilt models show that, if charge
stripes are present, the real
situation will be a mixture of LTO and LTT; 
or LTO, LTT and HTT tilts,
and not purely LTO or LTT as we have modeled.  The
presence of charged stripes, and regions between 
the stripes with no holes
present, implies that there coexists in the local 
structure regions with
strongly diminished tilts, and regions with 
large tilts (as much as 5$^\circ$)
even in the doped materials whose average tilt 
is 3$^\circ$ or less.
We would like to see whether this picture is 
consistent with our PDF
data.  We have made a simple test of this idea, 
and the results are shown
in Fig.~\ref{fig;fig11}. 
In this figure we compare PDFs which mimic the 
situation we would have if
there were a coexistence of $5^\circ$ tilts 
and $\approx 2^\circ$ tilts, with the
data from the $x=0.10$ sample whose average 
tilt is $3.6^\circ$.  We did this
by averaging the data-PDFs 
from the $x=0$ and $x=0.25$ samples. 
The former has
5$^\circ$ LTO tilts and the latter sample has tilts 
with less than $2^\circ$
of tilt amplitude.  It is clear from the figure that 
the incoherent mixture
of large and small tilts reproduces the data from 
the $x=0.10$ sample
excellently.  For comparison, the data-PDFs from 
the $x=0$ and $x=0.25$
data are reproduced in the lower panel of this figure.
There are large differences between them, yet when they 
are mixed together they
give an excellent account of the $x=0.10$ data.  This 
shows that the
data-PDFs for intermediate dopings are consistent with 
the presence of 
a distribution of local tilt magnitudes.  Interestingly, 
the largest 
fluctuations in the difference curve between the $x=0.10$ 
data and the 
mixture occur at $r\approx 2.8$~\AA\  and $r\approx 6$~\AA\ 
which is 
exactly where fluctuations are expected if local LTT-like 
tilts are present.  
This is also consistent with the fact that an inhomogeneous 
local tilt distribution exists in this material.  These results
are consistent with the observations of Hammel et 
al.~\cite{hamme;prb98,statt;prb95,hamme;prl93,marti;prl95} of an 
inhomogeneous tilt distribution in doped La$_{2-x}$Sr$_x$CuO$_4$ and
La$_2$CuO$_{4+\delta}$ from NMR measurements.  These authors measure the
electric field gradient at the La site which is very sensitive to
local tilt disorder and find a continuous distribution of La environments
similar to what would be expected from our models if local strains were 
also present.  We cannot compare quantitatively the tilt amplitudes because
there is not a direct relationship between EFG and octahedral tilt angle.
The octahedral tilt angle can be found from point-charge models, or 
more accurately from quantum cluster calculations~\cite{marti;prl95}; however, these authors
have not published values for the tilt amplitudes.  Nonetheless, there is  
excellent qualitative agreement.

\section{Conclusions}

In conclusion, we have shown that the PDF technique 
clearly differentiates 
between the presence and the absence of local 
tilts of CuO$_6$ octahedra. 
The magnitude of the local octahedral 
tilts in La$_{2-x}$Sr$_x$CuO$_4$ 
smoothly 
decreases as a function of Sr-content at T=10~K. 
This result is in agreement 
with the result for the average tilt angle 
magnitude,\cite{radae;prb94i,brade;pc94} 
obtained using 
standard crystallographic methods. Small but 
significant local tilts 
persist above $x=0.20$ in the HTT 
structural phase where the tilts are not allowed 
crystallographically. 

We showed that the PDF technique can
differentiate between different tilt directions, 
providing their magnitude is large enough. 
Our analysis showed that for the undoped material the local 
tilts clearly have the LTO 
symmetry, in agreement with crystallography, 
indicating that there
is no tilt disorder at this composition. For $x=0.05$, 
the local tilts
are still predominantly LTO-like.  As doping is increased, 
there is some
evidence that LTT-like tilts might be present, but in 
the interesting
superconducting region of the phase diagram the tilt amplitudes
are small enough that the PDF is at the limit of its sensitivity.

We have presented topological models for the tilt 
disorder we expect to be present in these
materials in the presence of charge-stripes.  
These imply that if charge
stripe domains exist locally there should be
a coexistence of large and small tilts, 
and LTO and LTT-like tilts locally.
A simple test using PDF data shows that this 
is consistent with our 
measured PDFs. A single crystal diffuse scattering, 
or electron diffraction
measurement may be useful for elucidating this point.

\acknowledgements
We are grateful to T. Egami, D. Haskel, 
D. Louca, R. McQueeney, E.A. Stern and 
J.D. Thompson  for stimulating discussions.
The work at Michigan State University 
was supported financially 
by NSF through grant DMR-9700966.  SJB is a 
Sloan Research Fellow of 
the Alfred P. Sloan Foundation. The work at 
Los Alamos was done under the 
auspices of the Department of Energy, 
under the contract W-7405-ENG-36. 
The experimental data were collected 
on the SEPD diffractometer at the IPNS 
at the Argonne National Laboratory. 
This facility is funded by the US 
Department of Energy under Contract W-31-109-ENG-38.


\begin{thebibliography}{10}
\bibitem{note1}
For example, the agreement factor, which is similar to the profile-weighted
  residuals function in Rietveld refinement~\cite{billi;b;lsfd98}, equals 0.194
  for the LTO model and 0.208 for the HTT model for the $x=0.30$ data set,
  where the LTO model has 13 independent variables and the HTT model has 11
  independent variables.
\end{thebibliography}
\end{document}